\documentclass[twocolumn,aps,prl,showpacs,amsmath,amssymb]{revtex4}

\usepackage{graphicx}
\usepackage{dcolumn}
\usepackage{bm}

\begin{document}

\title{Novel anisotropy in the superconducting gap structure 
of Bi$_2$Sr$_2$CaCu$_2$O$_{8+\delta}$ probed by 
quasiparticle heat transport}

\author{Yoichi Ando}
 \email{ando@criepi.denken.or.jp}
\author{J. Takeya}
\author{Yasushi Abe}
\author{X. F. Sun}
\author{A. N. Lavrov}
\affiliation{Central Research Institute of Electric Power Industry, 
Komae, Tokyo 201-8511, Japan.}

\date{\today}

\begin{abstract}
Since the nature of pairing interactions is manifested in the 
superconducting gap symmetry, the exact gap structure, particularly 
any deviation from the simple $d_{x^2-y^2}$ symmetry, would help 
elucidating the pairing mechanism in high-$T_c$ cuprates.  
Anisotropic heat transport measurement in 
Bi$_2$Sr$_2$CaCu$_2$O$_{8+\delta}$ reveals that the quasiparticle 
populations are different for the two nodal directions and thus the 
gap structure must be uniquely anisotropic, suggesting that pairing 
is governed by interactions with a rather complicated anisotropy. 
Intriguingly, it is found that the ``plateau" in the magnetic-field 
dependence of the thermal conductivity is observed only in the 
$b$-axis transport.
\end{abstract}


\maketitle

It has been established \cite{d-wave} that in the high-$T_c$ 
cuprates the superconducting gap $\Delta$ has essentially 
the $d_{x^2-y^2}$ symmetry, $\Delta(\mathbf{k}) = \Delta_0 \cos 2\phi$, 
which is depicted in Fig. 1 ($\Delta_0$ is the maximum gap and 
$\phi$ is the azimuthal angle).  
Although this information is very important, it does not help much 
in clarifying the high-$T_c$ mechanism, because all of the proposed 
theories of the high-$T_c$ superconductivity have successfully adopted 
the $d$-wave symmetry.  
It is thus important to obtain more detailed information on the 
gap structure, because any deviation from the simple $d_{x^2-y^2}$ 
symmetry would point to the role of a particular interaction 
in the paring.  
However, not much is known about the precise gap structure of 
the high-$T_c$ cuprates \cite{Mesot}, partly because the experimental 
techniques to selectively probe the nodal structures 
are limited \cite{note1}.

The most prominent feature of the $d$-wave gap is that 
it has four nodes where the gap magnitude vanishes.  
At finite temperature below $T_c$, quasiparticles (QPs) are 
always thermally excited near these nodes from the superconducting 
condensate; how many QPs are excited depends on the ``slope" of the 
gap at the node \cite{Lee}, because a steeper gap allows less 
phase space available below a given energy of $k_{B}T$.  
Since these QPs carry heat (in contrast to the Cooper pairs 
that do not carry heat), one can investigate the gap structure of 
unconventional superconductors by measuring the QP heat transport 
in the superconducting state \cite{Graf}.  
Moreover, because the QPs can only travel along the nodal 
directions \cite{Graf}, one can gain information about different 
nodes by tuning the direction of the heat current to the respective 
nodal directions; this makes it possible to study the detailed 
anisotropy of the gap structure in high-$T_c$ cuprates.  

\begin{figure}
\includegraphics[clip,width=6.5cm]{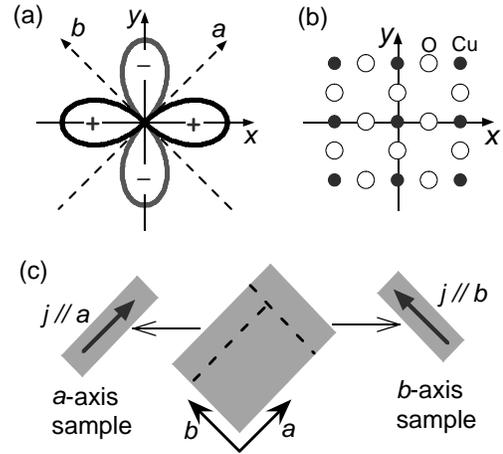}
\caption{\label{fig1}
Schematic pictures of the $d_{x^2-y^2}$ -wave gap and the 
present experiment.  
(a) Shape of the gap, $\Delta(\mathbf{k}) = \Delta_0 \cos 2\phi$, 
is plotted together with the directions of the $a$ and $b$ 
crystallographic axes of BSCCO. 
(b) Structure of the two-dimensional CuO$_2$ plane. 
(c) Two samples used for the measurements of the $a$- and $b$-axis 
transports are essentially the same crystal.}
\end{figure}

In recent years, the heat transport has been conveniently used to probe 
the gap structure of the high-$T_c$ cuprates.  
For example, it was reported \cite{Yu,Aubin} that the in-plane 
thermal conductivity showed a four-fold-symmetric change when the 
heat current was fixed along an antinodal direction and an 
applied magnetic field was 360$^{\circ}$ rotated in the CuO$_2$ plane, 
giving maxima for fields parallel to the nodal directions.  
This feature helped establishing the $d_{x^2-y^2}$ symmetry of the 
superconducting gap.  
The heat transport measurement was also employed in an intriguing 
proposal of a field-induced transition in the gap structure; 
namely, Krishana \textit{et al.} reported \cite{Krishana} that at low 
temperatures the magnetic-field ($H$) dependence of the thermal 
conductivity of Bi$_2$Sr$_2$CaCu$_2$O$_{8+\delta}$ (BSCCO) showed a 
``plateau" in high fields, and they proposed that this plateau region may 
correspond to a fully-gapped state where there are no QPs to carry heat. 
This experiment attracted a lot of attention because the field-induced 
phase transition in the superconducting state has a significant 
implication on the mechanism of the superconductivity; however, 
the origin of the plateau, as well as the reproducibility of the 
experimental result, are currently under intense debate 
(see Ref. \cite{Ando} and references therein).  
In this Letter, we report a striking in-plane anisotropy of the QP heat 
transport, which not only sorts out the ``plateau" issue but also 
suggests an unexpected anisotropy in the superconducting gap symmetry. 

\begin{figure}
\includegraphics[clip,width=5.5cm]{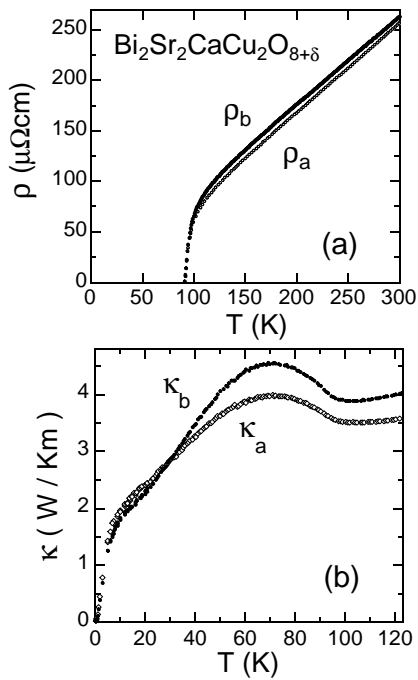}
\caption{\label{fig2}
Anisotropic transport in zero field.  
(a) Resistivity shows a very small, if any, in-plane anisotropy, 
indicating that the electronic properties are essentially isotropic 
in the normal state.  
(b) Thermal conductivity $\kappa$ shows a peculiar in-plane anisotropy 
below $T_c$ that suggests a difference in the QP populations along the 
two nodal directions.}
\end{figure}

We measure the thermal conductivity $\kappa$ of high-quality BSCCO 
single crystals using a conventional steady-state technique 
\cite{Ando,Takeya}.  
The crystals used here are grown by the floating-zone (FZ) technique 
\cite{Nakamura} and are annealed to be slightly underdoped 
(zero-resistance $T_c$ is 90.5 K); 
their crystallographic axes are determined by 
the X-ray Laue analysis within the accuracy of a few degrees.
To obtain reliable data on the in-plane anisotropy, we cut a 
piece of high-quality single-domain crystal into 
three parts and measure the $a$ and $b$ axis transports separately 
[Fig. 1(c)]; both the resistivity ($\rho_a$ and $\rho_b$) and the 
thermal conductivity ($\kappa_a$ and $\kappa_b$) are measured 
on the same samples.  
Note that in BSCCO the $a$ and $b$ crystallographic axes are 
along the nodal directions [Fig. 1(a)].  
The magnetic field is applied perpendicular to the $ab$ plane and 
the $H$-dependence of $\kappa$ is measured with the field-cooled 
procedure \cite{Ando} to avoid complications that are associated
with the vortex-pinning-related hysteresis \cite{Aubin2}.  

Figure 2 displays the temperature dependences of $\rho$ and $\kappa$ 
for the $a$ and $b$ directions; 
the resistivity shows only a small, if any, in-plane anisotropy, 
while the thermal conductivity demonstrates a non-trivial anisotropy 
that cannot be attributed to an error in the geometrical factors; 
namely, the peak height at $\sim$70 K is smaller along the $a$ axis 
and yet $\kappa_a$ becomes larger than $\kappa_b$ below $\sim$30 K. 
This anisotropy in $\kappa(T)$ suggests that the number of 
thermally-excited QPs is larger for the nodes along the $a$ axis, 
because the QP population near the nodes is more directly reflected 
in $\kappa$ at lower temperatures \cite{Hirschfeld}.  
(The anisotropy in $\kappa$ above $T_c$ is most likely due to phonons; 
it has been estimated \cite{Ando} that phonons are responsible for 
roughly $2/3$ of the total $\kappa$ at $T_c$.)
We note that the uncertainties in the absolute values in Fig. 2 are 
less than 10\% for both $\rho$ and $\kappa$.

\begin{figure}
\includegraphics[clip,width=5.5cm]{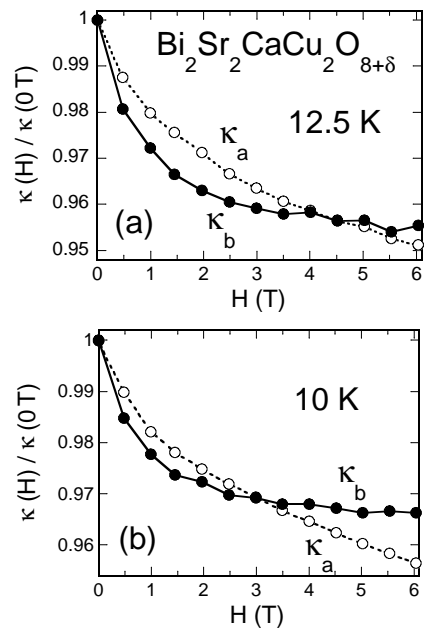}
\caption{\label{fig3}
Magnetic-field dependences of the anisotropic thermal 
conductivity at (a) 12.5 K and (b) 10 K.  The ``plateau" is observed 
only for the $b$ direction.  The data are normalized by the values 
in zero field. (At 12.5 K, $\kappa_a$ and $\kappa_b$ are 2.07 and 
1.96 W/Km; at 10 K, $\kappa_a$ and $\kappa_b$ are 1.96 and 1.81 W/Km).}
\end{figure}

The magnetic-field dependence of $\kappa$ reveals more striking 
in-plane anisotropy of the heat transport.  
Figure 3 shows the result of our precise $H$-dependence measurement 
(where the relative accuracy is $\sim$0.1\% \cite{Ando}) for 
$\kappa_a$ and $\kappa_b$ at 12.5 and 10 K up to 6 T.  
Interestingly, only $\kappa_b$ shows the plateau-like feature, 
indicating that there cannot be a true ``fully-gapped" state where 
all the nodes are lifted.  
(If a ``full gap" causes the QPs to disappear, there should be 
no QP heat transport in \textit{any} direction.)

\begin{figure}
\includegraphics[clip,width=8cm]{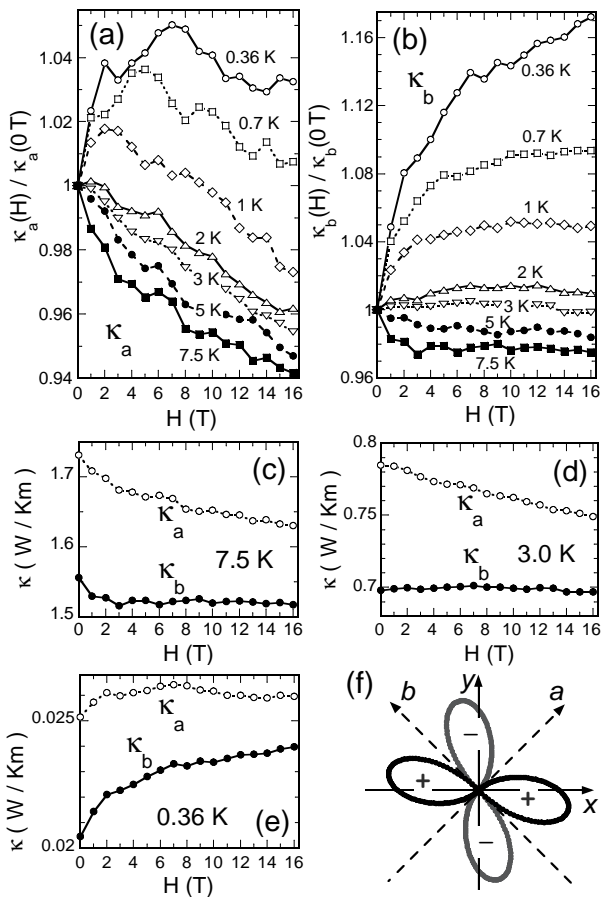}
\caption{\label{fig4}
Magnetic-field dependences of the anisotropic thermal 
conductivity at lower temperatures and in higher fields.  
(a,b) $H$-dependences of $\kappa_a$ and $\kappa_b$ at various 
temperatures down to 0.36 K.  
(c-e) Comparisons of $\kappa_a$ and $\kappa_b$ in absolute magnitude at 
representative temperatures.  
(f) Possible shape of the gap suggested by the data; 
the slope of the gap at the $b$-axis nodes is steeper than that 
at the $a$-axis nodes.}
\end{figure}

To further investigate the magnetic-field dependence of $\kappa$ 
down to lower temperatures and up to higher magnetic fields, 
we have measured $\kappa_a$ and $\kappa_b$ in a $^3$He refrigerator 
equipped with a 16 T magnet (with a bit worse signal-to-noise ratio).
Figures 4(a) and 4(b) show the $H$-dependences of $\kappa_a$ and 
$\kappa_b$, respectively, down to 0.36 K and up to 16 T; 
the trend is clear that $\kappa_a$ tends to decrease with $H$ while 
$\kappa_b$ tends to flatten or increase with $H$.  
Note that, since the phonons contribute to the heat transport 
in parallel to the QPs, the observed thermal conductivity is a sum of 
the QP term $\kappa_{\text{QP}}$  and the phonon term $\kappa_{ph}$.  
For BSCCO, it has been discussed \cite{Krishana,Ando,Aubin2} that 
more than 90\% of the total $\kappa$ is due to $\kappa_{ph}$ around 
10 K and this $\kappa_{ph}$ is independent of $H$; thus, although the 
$H$-dependence of the total $\kappa$ is rather weak in Figs. 4(a) and 
4(b), $\kappa_{\text{QP}}$ is actually changing a lot with $H$.

The physical origin of this anisotropic behavior can be inferred 
by plotting $\kappa_a$ and $\kappa_b$ at the same temperature together, 
taking the absolute values for the vertical axis [Figs. 4(c)-4(e)].  
In the simple kinetic theory, the QP thermal conductivity 
$\kappa_{\text{QP}}$ 
is proportional to both the number of QPs available for the 
heat transport, $N_{\text{QP}}$, and their mean free path 
$l_{\text{QP}}$; namely, 
$\kappa_{\text{QP}} \propto N_{\text{QP}} l_{\text{QP}}$.  
In $d$-wave superconductors, the quantized vortices produced by 
applied magnetic fields are expected to \textit{both} induce QPs 
near the nodes \textit{and} scatter them \cite{Kubert,Franz,Vekhter}.  
If there are already a large number of QPs along the $a$-axis in 
zero field, the increase in $N_{\text{QP}}$ due to the vortices may be 
inconsequential, while the vortex scattering will cause a noticeable 
reduction in $l_{\text{QP}}$; as a result, $\kappa_{\text{QP}}$ 
along the $a$-axis 
would tend to decrease with increasing $H$.  
On the other hand, if there are only a small number of QPs along the 
$b$-axis in zero field, both $N_{\text{QP}}$ and $l_{\text{QP}}$ 
along the $b$-axis 
will be noticeably changed due to the vortices and the changes in 
$N_{\text{QP}}$ and $l_{\text{QP}}$ will tend to cancel each other, 
leading to a weaker $H$-dependence along the $b$-axis.  
(Actually, one of the probable explanations of the ``plateau" in 
$\kappa(H)$ is that it comes from an exact cancellation of the 
two effects \cite{Franz}.]) 
This phenomenology explains the behaviors of $\kappa_{a}(H)$ and 
$\kappa_{b}(H)$ in Figs. 4(c) and 4(d) (7.5 and 3.0 K) and is 
consistent with their relative magnitude.  
At lower temperature [Fig. 4(e)], it appears that the changes in 
$N_{\text{QP}}$ dominate over the changes in $l_{\text{QP}}$, 
causing $\kappa$ to 
increase with $H$.
(Unfortunately, more quantitative analysis of the data is hindered 
by the lack of proper understanding of the vortex scattering of 
the QPs in the cuprates, which remains rather controversial 
\cite{Franz,Vekhter}.)

Thus, the novel anisotropies in both the temperature dependence and 
the magnetic-field dependence of the thermal conductivity 
consistently tell us that there are more QPs near the nodes along the 
$a$-axis; this gives strong evidence that the gap structure notably 
deviates from the $d_{x^2-y^2}$ symmetry.  
Mixing of a small $s$-wave component, which the orthorhombic distortion 
may cause \cite{s-mix}, would \textit{not} produce the observed 
anisotropy.  
One simple possibility to make the gap consistent with the observed 
anisotropy is to introduce ``higher harmonics"; for example, for a gap 
with an expression 
$\Delta(\mathbf{k}) = \Delta_0 \cos 2\phi - \Delta_1 \sin 4\phi$, 
the slope at the nodes is gentle along the $a$ axis and steeper along 
the $b$ axis [Fig. 4(f)] \cite{note3}.  
Of course, some non-analytic modulation of the $d_{x^2-y^2}$ symmetry 
may also be possible.  

Since the reproducibility of the plateau in $\kappa(H)$ has been a 
matter of debate \cite{Ando}, we comment on the reproducibility issue. 
We reported previously that the plateau was not observed in the 
majority of samples \cite{Ando}; recently, we found that this 
majority of samples were cut along the $a$-axis, because our 
crystals usually grow along the $a$-axis in the FZ furnace.  In fact, 
the $\kappa(H)$ data published in Ref. \cite{Ando} agree with the 
$\kappa_a$ data reported here.  To check whether our salient observation 
reported here is reproducible, we have measured another $b$-axis sample 
(Sample B2) that was taken from a different batch where the crystal 
grew along the $b$-axis.
We have also measured a sample (Sample M) which was cut along the direction 
45$^{\circ}$ rotated from the $a$-axis (Cu-O-Cu bond direction). 
As shown in Fig. 5, Sample B2 (which is optimally-doped with $T_c = 95$ K) 
reproduces the behavior of the first $b$-axis sample [Fig. 4(b)], 
and Sample M shows the behavior which is essentially an average of 
$\kappa_a(H)$ and $\kappa_b(H)$.  
These results give confidence in the reproducibility as well as the 
intrinsic nature of our observation.

\begin{figure}
\includegraphics[clip,width=8cm]{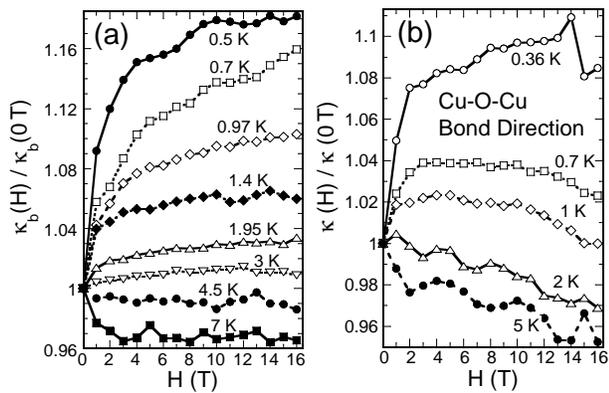}
\caption{\label{fig5}
$H$ dependences of $\kappa$ in (a) the second $b$-axis sample (Sample B2) 
and (b) a sample cut along the Cu-O-Cu bond direction (Sample M).}
\end{figure}

It is useful to note that the angle-resolved photoemission experiments 
have reported almost completely four-fold-symmetric intrinsic Fermi 
surface for BSCCO \cite{Shen}, and therefore the novel anisotropy of the 
gap found here is not likely to be due to a Fermi surface anisotropy 
\cite{note4}. 
(In fact, the normal-state resistivity shows negligibly small in-plane 
anisotropy [Fig. 2(a)], which is consistent with the almost-isotropic 
Fermi surface.)
Therefore, the striking anisotropy in the gap structure suggests 
that the driving force of the gapping, or the ``glue" to bind the 
electrons into Cooper pairs, is anisotropic itself. 
From this point of view, it would be difficult to imagine that the 
antiferromagnetic interaction on the nearly-square CuO$_2$ planes is 
wholly responsible for the pairing, though it is often considered to be 
the most likely candidate.  
On the other hand, the phonons are expected to be anisotropic in 
BSCCO because of the pronounced modulation structure \cite{incomm}, 
and thus the role played by the phonons in pairing is an 
interesting possibility.  
Another possibility is a role of the charge inhomogeneity \cite{Pan}, 
which could be related to the isotropic or nematic phases of the 
electronic liquid crystal \cite{Kivelson,stripe}, though the impact of 
such inhomogeneous structure on the gapping and on the QP behavior is 
not understood yet.  
In any case, the novel anisotropy in the gap structure suggested by the 
present experiment should help clarifying how the superconducting gap is 
formed in high-$T_c$ superconductors.

In summary, we have measured the thermal conductivity $\kappa$ of BSCCO 
along the $a$ and $b$ axes in essentially the same high-quality 
single crystal down to 0.36 K.  
It is found that the temperature dependence \textit{and} 
the magnetic-field dependence of $\kappa$ both show peculiar in-plane 
anisotropy, and the observed anisotropy is consistent with the 
interpretation that there are more QPs near the nodes along the $a$ axis.  
This result gives evidence that the $d$-wave gap is uniquely distorted
in BSCCO, which suggests that the electron pairing is contributed by the 
interaction that develops a pronounced anisotropy 
between the $a$ and $b$ axes. 
In addition, our measurement clarifies that only $\kappa_b$ shows the 
``plateau" in its magnetic-field dependence; this new information should 
help elucidating the true origin of this intriguing plateau.

We thank D. S. Dessau, A. Fujimori, P. J. Hirschfeld, A. Kapitulnik, 
Z. X. Shen, M. Sigrist, and I. Vekhter for helpful discussions.  
Y. Abe acknowledges support from JSPS and X.F.S. acknowledges 
support from JISTEC.


\begin{thebibliography}{}
\bibitem{note1}
In principle, the angle-resolved photoemission should be able to 
resolve the $\mathbf{k}$ dependence of the gap around the Fermi surface; 
however, in BSCCO, measurements along the $a$ direction is complicated 
due to the existence of the shadow bands.
\bibitem{note3}
The component ``$\sin 4\phi$" would not occur in a simple expansion of 
the gap function in Fermi surface harmonics and thus is highly unusual; 
however, the most natural harmonic ``$\cos 6\phi$" \cite{Mesot}, which 
is consistent with the $d_{x^2-y^2}$ symmetry, does not produce the 
observed anisotropy.
\bibitem{note4}
The ``shadow" Fermi surfaces, which show up anisotropically, are caused 
by the extrinsic effect of the modulations in the Bi-O layers 
\cite{incomm} and thus are not a sign of a true Fermi surface anisotropy.
\end{thebibliography}
\end{document}